# The binary central star of the bipolar pre-planetary nebula IRAS 08005−2356 (V510 Pup)

Rajeev Manick[1,2], Brent Miszalski[3], Devika Kamath[4,5], Patricia A. Whitelock[1,9], Hans Van Winckel[6], Bruce J. Hrivnak[7], Brad N. Barlow[8], Shazrene Mohamed[1,9,10]

[1] *South African Astronomical Observatory, PO Box 9, Observatory, 7935, South Africa*
[2] *Univ. Grenoble Alpes, CNRS, IPAG, 38000 Grenoble, France*
[3] *Australian Astronomical Optics - Macquarie, Faculty of Science and Engineering, Macquarie University, North Ryde, NSW 2113, Australia*
[4] *Department of Physics & Astronomy, Macquarie University, Sydney, NSW 2109, Australia*
[5] *Astronomy, Astrophysics and Astrophotonics Research Centre, Macquarie University, Sydney, NSW 2109, Australia*
[6] *Instituut voor Sterrenkunde, K.U.Leuven, Celestijnenlaan 200D bus 2401, B-3001 Leuven, Belgium*
[7] *Department of Physics and Astronomy, Valparaiso University, Valparaiso, IN 46383, USA*
[8] *Department of Physics and Astronomy, High Point University, High Point, NC 27268, USA*
[9] *Department of Astronomy, University of Cape Town, Private Bag X3, Rondebosch 7701, South Africa*
[10] *National Institute for Theoretical Physics (NITheP), KwaZulu-Natal, South Africa*
E-mail:rajeev.manick@univ-grenoble-alpes.fr

**ABSTRACT**
Current models predict that binary interactions are a major ingredient for the formation of bipolar planetary nebulae (PNe) and pre-planetary nebulae (PPNe). Despite years of radial velocity (RV) monitoring, the paucity of known binaries amongst the latter systems is insufficient to examine this relationship in detail. In this paper, we report on the discovery of a long period (P=2654 ± 124 d) binary at the centre of the Galactic bipolar PPN, IRAS 08005−2356 (V510 Pup) determined from long-term spectroscopic and near-infrared time series data. The spectroscopic orbit is fit with an eccentricity of 0.36 ± 0.05 that is similar to other long period post-AGB binaries. Time resolved H$\alpha$ profiles reveal high-velocity outflows (jets) with de-projected velocities up to $231^{+31}_{-27}$ km s$^{-1}$ seen at phases when the luminous primary is behind the jet. The outflow traced by H$\alpha$ is likely produced via accretion onto a main sequence companion for which we calculate a mass of 0.63 ± 0.13 M$_\odot$. This discovery is one of the first cases of a confirmed binary PPN and demonstrates the importance of high-resolution spectroscopic monitoring surveys on large telescopes in revealing binarity among these systems.

## 1 INTRODUCTION

After losing most of their outer envelope on the Asymptotic Giant Branch (AGB), low- to intermediate-mass stars evolve through the post-AGB phase. During this post-AGB evolution, some stars go through a PPN phase which is characterised by systems that display a well-resolved nebula, but whose central star has not evolved to be hot enough to emit a sufficient quantity of Lyman continuum photons to ionize the surrounding remnant of the AGB envelope. Central stars of PPNe are generally luminosity class I objects with F or G spectral types. They display strong infrared excesses at long wavelengths (e.g Hrivnak et al. 1989) and CO emission (e.g Bujarrabal et al. 2001) indicative of residual AGB envelope material that is detached from the stellar photosphere.

Many PPNe are also a subset of optically bright post-AGB stars with dust excess (Van Winckel 2003), which can themselves be divided into two categories (shell or disk sources) based on their spectral energy distribution (SED, Gezer et al. 2015). The shell sources display a double-peaked SED indicating the presence of cool dust (∼ 150 K) expanding in a shell detached from the post-AGB core. Most of these show evidence of their nebulosity in reflected starlight (Ueta et al. 2000; Sahai et al. 2007; Siódmiak et al. 2008).

The disk sources show a near-infrared excess which originates from warm dust with a range of temperatures, starting at the dust sublimation temperature. These stars are found to be binaries with periods in the range 100 to 3000 days (Van Winckel et al. 2009; Oomen et al. 2018) and are surrounded by circumstellar material in the form of a stable Keplerian circumbinary disk (De Ruyter et al. 2006; Kluska et al. 2018). Studies by Gezer et al. (2015), Oomen et al. (2019) and Kamath & Van Winckel (2019) show that there is a close correlation between the presence of such a disk and the chemical depletion process (Waters et al. 1992), suggesting an interaction between the circumbinary disk and the central source in post-AGB binaries. Most of these disk post-AGB sources do not show a resolved nebulosity. There are, however, some examples of disc post-AGB binaries, such as the Red Rectangle (HD 44179, Waelkens et al. 1996), 89 Her (Waters et al. 1993; Bujarrabal et al. 2013; Hillen et al. 2013) and IW Car (Bujarrabal et al. 2017), that display a resolved nebulosity either in the optical or in CO and have been identified as PPNe in the literature, but are significantly less massive analogues of shell PPNe.

Past PPNe imaging surveys at optical or near-IR wavelengths (Ueta et al. 2000; Sahai et al. 2007; Siódmiak et al. 2008) revealed a wide range of aspherical nebular structures many of which are bipolar. Such bipolar PPNe are often found to be accompanied by a dense, large-scale equatorial dusty torus/waist (Sahai et al. 2007; Huggins 2007) and high velocity bipolar outflows as in GL 618, (Trammell & Goodrich 2002), IRAS 16594−4656, (Volk et al. 2006) and IRAS 16342−3814, (Sahai et al. 2017). The exact origin of these outflows is still not well understood, although they are believed to be launched due to accretion by a companion in a binary system (Soker 2002; Nordhaus & Blackman 2006; Balick et al. 2013; Akashi & Soker 2013; García-Segura et al. 2020).

To search for binaries in PPNe, Hrivnak et al. 2011, 2017 conducted radial velocity (RV) monitoring of several of the brightest aspherical PPN shell sources, but surprisingly found no clear binaries. Whether this paucity in binary detection in shell PPNe is due to binary evolutionary channels being intrinsically rare amongst PPNe or rather due to observational biases remains a puzzle.

As part of a wider program to find binaries among central stars of planetary nebulae (CSPNe) and PPNe (e.g. Miszalski et al. 2018a,b, 2019), we monitored IRAS 08005−2356 using high resolution spectroscopy on the Southern African Large Telescope (SALT). We included this object in our RV monitoring program since it was bright and a suspected binary candidate based on three distinct features: 1) observations of dual dust chemistry: the presence of both C-rich and O-rich material in the circumstellar envelope (Hrivnak 1995; Sánchez Contreras et al. 2008), 2) the bipolar nature of the outflows accompanied by a dusty disk/torus (Ueta et al. 2000; Sahai & Patel 2015) and 3) a long term periodic variability seen in the photometric time series (this work). Despite these and other strong signs of binarity in the nucleus of this object (e.g Trammell et al. 1994; Bakker et al. 1996), its binary nature remained unproven spectroscopically.

In this paper we report on the long period binary nature of the central star of the PPN IRAS 08005−2356. In Section 2 we outline the main properties of IRAS 08005−2356. We then describe our observations and data analysis in Section 3. In Section 4 we discuss and conclude the main results presented in this paper.

## 2 TARGET DETAILS

IRAS 08005−2356 was first observed by Slijkhuis et al. (1991) as part of an infrared (IR) survey of cool IRAS sources to look for PPNe. They identified the bright (V = 11.5) central star, also known as V510 Pup, and classified it as an F5 spectral-type supergiant.

Based on spectropolarimetric observations, Trammell et al. (1994) first found evidence for polarisation due to bipolar outflows in which the two lobes act as a reflection nebula with the central star obscured by a dense dusty torus. Slijkhuis et al. (1991) computed a total extinction ($A_v$) of 3.4 mag towards the central star. Indirect evidence for high velocity outflows were provided by Bakker et al. (1997) who noted the presence of several narrow, double-peaked emission lines in the spectrum, which they associate with an accretion disk around a companion launching outflows. The first evidence of the outflows likely being bipolar was obtained by Ueta et al. (2000) who resolved the bipolar reflection nebula with the *Hubble Space Telescope*. They determined the size of the nebula to be ∼ 2.68″ ×1.42″. Outflows have also been resolved in OH data (Zijlstra 2001) and in CO and SiO by Sahai & Patel (2015), who reported CO outflow velocities reaching up to 200 km s$^{-1}$.

Detailed radiative transfer spectropolarimetric modeling was done by Oppenheimer et al. (2005) which included dust-scattering and absorption at optical wavelengths. Among the main results from their model were 1) a high mass-loss rate of $4.4 \times 10^{-6}$ M$_\odot$ yr$^{-1}$ and 2) the presence of micrometre sized dust grains in the circumstellar material. Their model constrained the disk inclination (the angle between the perpendicular to the disk and the observer's line-of-sight) to be 60 ±5 ° and included both warm and cold dust at ∼24 au and ∼2740 au, respectively (Oppenheimer et al. 2005).

## 3 OBSERVATIONS AND DATA ANALYSIS

### 3.1 Spectroscopic data

We obtained 23 observations of IRAS 08005−2356 from 2017 to 2021 using SALT High-resolution spectroscopy (HRS, Bramall et al. 2010, 2012; Crause et al. 2014) which is a dual-beam, fibre-fed Échelle spectrograph (Buckley et al. 2006; O'Donoghue et al. 2006). We used the medium resolution (MR) mode of HRS covering 3800-8900 Å with resolving powers (R = $\lambda/\triangle\lambda$) of 43000 and 40000 for the blue and red arms, respectively. This results in a velocity resolution of ∼ 0.2 km s$^{-1}$ at a wavelength of 7771 Å with the MR mode fibre aperture of 1.6″. During an observation a second fibre, separated at least 20″ from the science fibre, simultaneously observed the sky spectrum. Regular bias, ThAr arc lamp and quartz lamp flat-field calibrations are taken as part of SALT operations. These observations were carried out under programmes with IDs 2017-1-MLT-010 (PI: Miszalski), 2018-2-MLT-007 (PI: Miszalski) and 2019-2-SCI-031 (PI: Manick).

Basic reduction of the SALT raw data was done using the pipeline of Crawford et al. (2010) and then reduced using the MIDAS pipeline of Kniazev et al. (2016) that is based on the ECHELLE (Ballester 1992) and FEROS (Stahl et al. 1999) packages. Heliocentric corrections were applied to the RVs using the VELSET task of the RVSAO package (Kurtz & Mink 1998).

We gathered an additional 32 medium-resolution long slit spectra (R ∼ 27500) from the Small and Medium Aperture Research Telescope System (SMARTS) 1.5 m telescope at the Cerro Tololo Inter-American Observatory (CTIO). These were recorded between 2014 and 2018 using the fiber-fed CHIRON Échelle spectrometer (Tokovinin et al. 2013) which collects light through a 2.7″ on-sky aperture. Extracted and wavelength-calibrated spectra were delivered by the pipeline running at Yale University (Brewer et al. 2014). The final velocity resolution achieved in our CTIO observations were ∼ 3.8 km s$^{-1}$ at a wavelength of 7771 Å.

Another two long slit Échelle spectra were obtained from the William Herschel Telescope (WHT) and the Very Large Telescope (VLT). The WHT observations were performed on 1st March 1994 using the Utrecht Échelle Spectrograph (UES, R = 110000), which has an on-sky aperture size of 1.1″. The two-dimensional WHT spectra were reduced following the standard procedure for Échelle spectroscopy using IRAF astronomical routines. The radial velocity resolution achieved in our spectra was ∼ 1.4 km s$^{-1}$ at a wavelength of 7771 Å.

The order merging for the WHT data was not optimal (e.g. Mazeh et al. 1997) which affected some lines in the spectrum that were omitted in our analysis. The VLT spectrum was taken on 6 April 2001 using UVES (R=40000) and was available publicly (ESO programme ID:67.D-0243, PI: Van de Steene). The on-sky aperture of UVES is 1.0″ and the velocity resolution achieved in the VLT spectrum is 0.9 km s$^{-1}$ at 7771 Å.

**Table 1.** Observation log of IRAS 08005−2356. Column 1 represents the mid-point HJD of each exposure with duration given in column 2. Column 3 shows the weighted mean of the Heliocentric RVs obtained from the OI triplet (see Section 3.4) and column 4 is the standard deviation from the mean. The orbital phase is listed in column 5. The full RV dataset can be accessed online through Vizier.

| HJD (d) | Exptime (sec) | RV (km s$^{-1}$) | $\sigma$ (km s$^{-1}$) | Phase | Telescope |
|---|---|---|---|---|---|
| 2449412.51 | 900 | 73.3 | 0.8 | 0.20 | WHT |
| 2452005.57 | 1980 | 67.7 | 0.8 | 0.28 | VLT UVES |
| 2457010.82 | 1200 | 65.6 | 1.5 | 0.20 | CTIO 1.5m |
| 2457051.76 | 1200 | 70.0 | 2.2 | 0.22 | CTIO 1.5m |
| 2457472.60 | 1200 | 73.9 | 1.3 | 0.38 | CTIO 1.5m |
| 2458142.72 | 1200 | 77.3 | 1.9 | 0.64 | CTIO 1.5m |
| 2458505.56 | 600 | 78.9 | 1.1 | 0.77 | SALT |
| 2458825.44 | 600 | 66.7 | 0.5 | 0.90 | SALT |
| 2458990.23 | 600 | 64.3 | 0.5 | 0.96 | SALT |
| 2459209.39 | 800 | 63.6 | 0.4 | 0.04 | SALT |

**Table 2.** The *JHKL* photometry for IRAS 08005−2356. The uncertainty on the photometry is 0.02 mag at *JHK* and 0.05 mag at *L*. The full table is available online.

| HJD (d) | J | H | K | L |
|---|---|---|---|---|
|  |  | (mag) |  |  |
| 2446465.49 | 8.21 | 7.04 | 5.75 | 3.90 |
| 2447920.42 | 8.06 | 6.84 | 5.53 | 3.74 |
| 2447962.41 | 8.03 | 6.81 | 5.51 | 3.72 |
| 2448028.22 | 8.06 | 6.84 | 5.54 | 3.78 |

A subset of the log of our spectroscopic observations is shown in Table 1. The full table is available online.

### 3.2 Photometric data

Near-infrared *JHKL* photometry (Table 2) was obtained from the South African Astronomical Observatory (SAAO) as part of program investigating variability in IRAS sources (e.g. Whitelock & Feast 1984; Menzies & Whitelock 1988). Regular monitoring was performed for 16 years from 1990 to 2005 with the MkII infrared photometer on the 0.75m telescope at SAAO Sutherland. This uses a single channel InSb detector and a 34″ aperture, which encloses the resolved nebula in its entirety. The final isolated observation was obtained in 2010 also on the 0.75 m, about 24 years after the first. The photometry, listed in Table 2, is on the SAAO system as described by Carter (1990) and the uncertainty in the *JHK* and *L* bands is 0.02 mag and 0.05 mag, respectively.

We also include V-band photometry from 2010-2018 that was obtained as part of a program to monitor the light variability of PPN candidates using the telescopes of the Southeastern Association for Research in Astronomy (SARA) at CTIO and Kitt Peak National Observatory (KPNO) (e.g., Hrivnak et al. 2020a,b). These were combined with *V*-band data from the All-Sky Automated Survey for Supernova database (Kochanek et al. 2017) from 2016-2018 and *V*-band data from the earlier All Sky Automated Survey, version 3[1] from 2000-2009. Details of this study will be published elsewhere (Hrivnak et al. 2021, in preparation). The resulting V light curve is shown in the top panel in Figure 1.

### 3.3 Photometric analysis

The *JHKL* data were analysed using a multi-band Fourier analysis VanderPlas & Ivezić (2015), but omitting the first and last observations (because they have large time gaps from the bulk of the observation) which yielded a period of 2741 d., which yielded a period of 2741 d. The light-curves show a very clear periodicity at all wavelengths with amplitudes (half peak-to-peak) of 0.057 ± 0.007, 0.111 ± 0.009, 0.151 ± 0.006 and 0.149 ± 0.007 mag at *J*, *H*, *K* and *L* respectively. The *J* curve is distinctly erratic, while the *H* curve seems more regular, but the minimum is flat bottomed, suggestive of a significant contribution from a non-variable source. The *K* and *L* data alone yield a period of 2727 ± 26 days, which we use in combination with the RV period to derive the orbital period of the system (see Section 3.4). Figure 1 shows all of the near-IR observations, phased on the orbital period (see Section 3.4).

We hypothesize that the modulation which is most clearly seen at *L* (3.5$\mu$m) is the result of the eccentric orbits of the stars causing differential heating of dust in the circumbinary disk/torus (see Oppenheimer et al. 2005). If dust were the only contributor to the measured magnitudes, then at shorter wavelengths the star would be fainter and the amplitudes larger than we actually observe. It seems likely therefore that scattered light from the nebula makes an increasing contribution to the observed flux as we move to shorter wavelengths. Radiative transfer modelling is required to investigate this further and is outside the scope of this work. We also note that the RV, near-IR and visual observations are not contemporaneous, and any errors in our derived period will affect the relative phasing of the data. There is some evidence for long term changes in the photometry, e.g., the last *JHKL* observation, obtained 1775 days after the previous one, (in blue on Figure 1) does not fit well with the earlier observations.

The *V*-band photometric time series data phased on the orbital period (see Section 3.4) is shown for comparison in Figure 1. It has an amplitude of 0.083 ± 0.004 mag. In addition to the long term variability, it shows short term variability (scatter) most likely related to pulsational variations. The analysis of the short-term photometric variability is outside the scope of this paper since we are concerned here with the long term behaviour to corroborate the spectroscopic observations.

### 3.4 Orbital analysis

The RVs were computed by fitting Voigt profiles (see Appendix A) to deduce accurate centers of OI absorption lines at 7771.94 and 7774.17 Å using the python LMFIT module (Newville et al. 2016). In our analysis, we excluded the third line at 7775.39 Å in the OI triplet as these were not clearly visible (likely blended) in some of the CHIRON spectra. A weighted mean RV was computed by combining the mean RV of each of the two lines. The errors were deduced by computing the standard deviation of the RV deduced from each line. A subset of the computed RVs are presented in Table 1.

The orbital period of the system was derived using a combined analysis of the RV time series, *K* and *L* band data. We used the method of VanderPlas & Ivezić 2015 to perform a multi-band Lomb Scargle analysis on the three datasets, which yielded a period of 2654 ± 124 d, which we adopt as the orbital period of the system. Both the photometric time series Figure 1 and the RV time series Figure 2 are phased on this period.

---
[1] http://www.astrouw.edu.pl/asas/

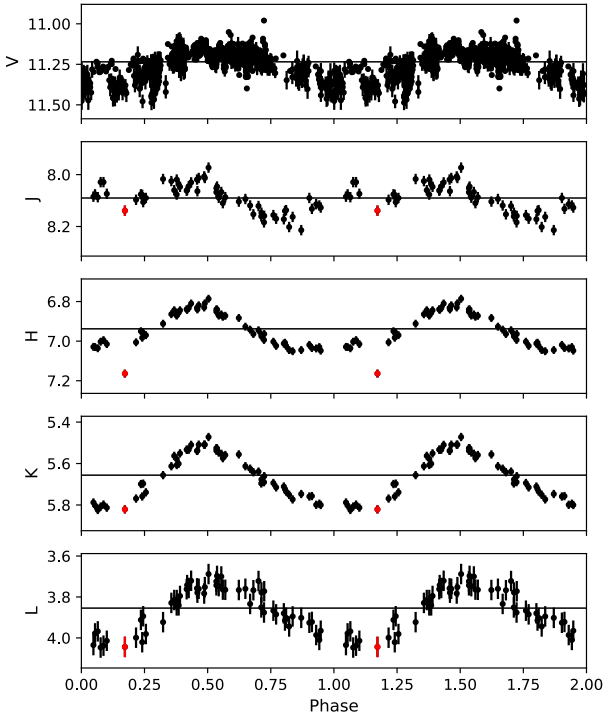

**Figure 1.** The *V J H K L*-band light curves of IRAS 08005−2356 phased on a period of 2654 d (see section 3.4). The red point is the last in the time series and is faintest in *H*.

This period was fixed in the Keplerian model fit, allowing all other parameters to vary. The fitted Keplerian model results in an eccentricity of 0.36 ± 0.05 which we find to be significantly different from zero at a 5-$\sigma$ level according to the Lucy-Sweeney test (Lucy & Sweeney 1971). The uncertainties on the orbital parameters were computed using Monte Carlo simulations on the RV data following the same method as described by Miszalski et al. (2018a). The final error on the orbital period was combined in quadrature using the uncertainty in the RV, *K* and *L* band periods. Table 3 presents the orbital parameters derived from the Keplerian orbit fit and the derived uncertainties for each orbital parameter.

Adopting a mean luminosity of 6640 ± 340 L$_\odot$ (Oppenheimer et al. 2005; Sahai & Patel 2015) and a mean effective temperature of 6750 ± 250 K (Slijkhuis et al. 1991; Oppenheimer et al. 2005), we determine the current mass of the primary to be $M_1 \approx 0.57 \pm 0.1$ M$_\odot$ using the post-AGB luminosity-core mass relation of Miller Bertolami (2016). Assuming this mass for the primary and assuming that the orbital inclination is the same as the disk inclination of 60 ± 5 degrees derived by (Oppenheimer et al. 2005), we derive a companion mass of 0.63 ± 0.13 M$_\odot$. We note that since most of the light is dominated by scattering (Oppenheimer et al. 2005), this would have an effect on the determination of the orbital inclination and hence the mass of the companion.

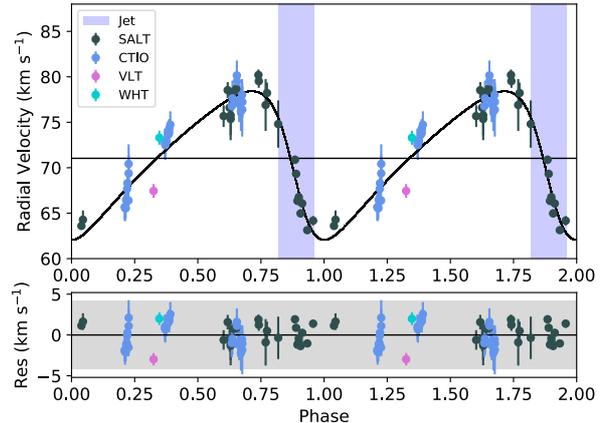

**Figure 2.** A plot of the spectroscopic RV data (computed using a heliocentric rest-frame) phased with the orbital period of 2654 ± 124 days. Top panel: a Keplerian model (black curve) is fit to the data with an eccentricity of 0.36 ± 0.05. The horizontal line is the systemic velocity at 71.0 ± 0.5 km s$^{-1}$. The blue shaded region indicates the phase range where the jets appear. Bottom panel: Residuals to the fit. The grey shaded region represents a standard deviation of 3-$\sigma$.

**Table 3.** Orbital parameters derived from the Keplerian model fit by fixing the orbital period at 2654 d (see Section 3.4). The companion mass is derived assuming a primary mass of 0.57 ± 0.1 M$_\odot$ and an orbital inclination of 60 ± 5 degrees based on the results of Oppenheimer et al. (2005).

| Parameter | Value | $\sigma$ |
| --- | --- | --- |
| P (d) | 2654 | 124 |
| e | 0.36 | 0.05 |
| $K_1$ (km s$^{-1}$) | 8.2 | 0.4 |
| $T_0$ (HJD) | 2459105 | 24 |
| $\gamma$ (km s$^{-1}$) | 71.0 | 0.5 |
| $\omega$ (°) | 105 | 65 |
| $a_1 \sin i$ (au) | 1.9 | 0.2 |
| i (°) | 60 | 5 |
| f(m) (M$_\odot$) | 0.12 | 0.02 |
| $M_1$ (M$_\odot$) | 0.57 | 0.10 |
| $M_2$ (M$_\odot$) | 0.63 | 0.13 |

### 3.5 Time series behaviour of H$\alpha$

Figure 3 shows dynamic plots of the H$\alpha$ and OI lines when they are phased on the orbital period. The intrinsic H$\alpha$ profile (which shows a double-peaked emission with a central absorption component) gains an extended blue-shifted absorption feature between phases ∼ 0.82 and 0.96, where phase zero corresponds to the minimum of the RV curve.

Similar phenomena in the H$\alpha$ profile are seen among analogous post-AGB binaries (e.g Gorlova et al. 2015; Bollen et al. 2017, 2019), which are indicative of a high-velocity outflow (jets) originating from the companion. The jets are seen in absorption when the luminous primary is behind the jet-launching companion since the jet scatters continuum photons of the primary out of our line-of-sight. We sim-ilarly interpret these absorption features in IRAS 08005−2356 as indicative of a high-velocity outflow from the companion.

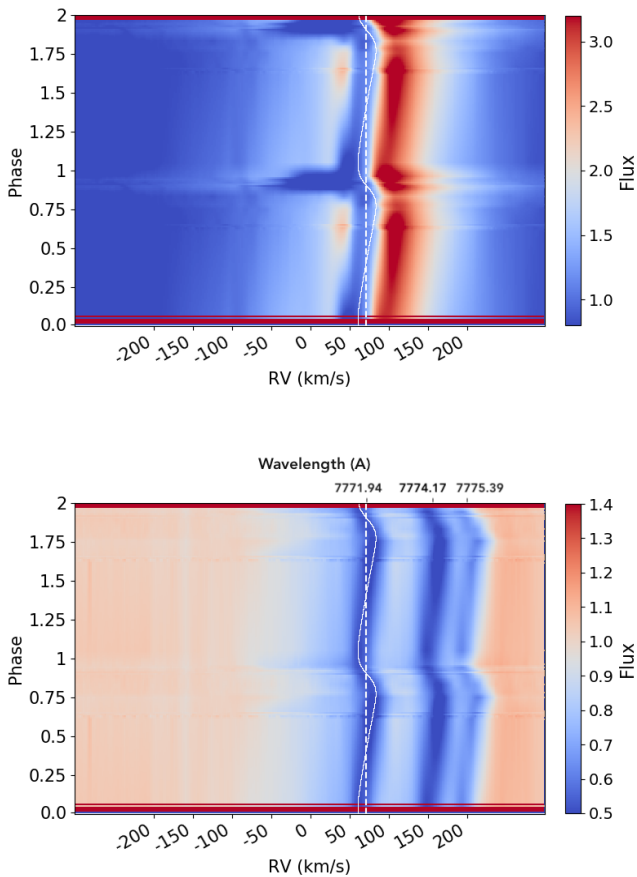

**Figure 3.** Top panel: Dynamic plot of H$\alpha$ profiles (normalised to the continuum level) phased on the orbital period of 2654 ± 124 days. Phase zero corresponds to the minimum of the Keplerian orbit fit (Figure 2). Between phases ∼ 0.82 and 0.96, the H$\alpha$ profile display extended absorption features (in blue) indicating that we are looking through the outflow (jet) launched by the companion. Bottom panel: A dynamic plot showing the OI lines (normalised to the continuum level) phased on the orbital period. The orbital fit is also shown on both plots as the white curve and the vertical dashed line is at a systemic velocity of 71 km s$^{-1}$. The rest wavelengths of the OI lines are shown on the top axis.

Our measurement of the maximum width of the H$\alpha$ absorption profile occurring at phase 0.9 gives a value of ∼ 200 ± 15 km s$^{-1}$. Assuming an orbital inclination of 60 ± 5 degrees, this yields a de-projected velocity of $231^{+31}_{-27}$ km s$^{-1}$, in agreement with the results of Sahai & Patel (2015). This outflow velocity corresponds to the escape velocity of a main sequence star, rather than that of a compact object (Sánchez Contreras et al. 2008) and is consistent with the companion mass we derive from the orbital analysis (Table 3).

## 4 DISCUSSION AND CONCLUSIONS

In this study we have proven the binary nature of the central star of the PPN IRAS 08005−2356. With an orbital period of 2654 ± 124 d, this is currently the longest measured orbital period within a PPN. We fit the RV orbit with an eccentricity of 0.36 ± 0.05. Adopting a primary mass of $M_1$ = 0.57 ± 0.1 $M_\odot$ and an orbital inclination of 60±5 degrees (Oppenheimer et al. 2005), we derive the companion mass to be 0.63 ± 0.13 $M_\odot$.

We find evidence of high velocity outflows (jets) in the H$\alpha$ profiles which are highly dependent on the orbital phase. Analysis of the H$\alpha$ profile shows that the jets can reach de-projected velocities of up to ∼ $231^{+31}_{-27}$ km s$^{-1}$, consistent with previous results for other post-AGB binaries. We also find evidence that the jets are being launched by the companion, since the jets only appear at phases close to superior conjunction, when the jet-launching companion is between the observer and the evolved star.

Our results suggest that binary interaction plays a key role in producing bipolar systems like IRAS 08005−2356. The jets may be supplying the primary force in shaping the surrounding nebula into a bipolar structure. Our results also complement models (e.g. Nordhaus & Blackman 2006; Raga et al. 2009) and PNe (e.g. Boffin et al. 2012) that require a binary interaction to produce such bipolar outflows.

Although, previous studies showed that measured orbits among PPNe are rare, our analysis of IRAS 08005−2356 suggests there may be more binaries to be found among such systems. High-resolution spectroscopic monitoring surveys on large telescopes have the potential to detect the orbits of such long period systems, and those PPNe with long period light variations are among the most likely binary candidates. Further studies aimed at finding binaries among PPNe will help bridge important gaps in our understanding of binary evolution among evolved stars.

## ACKNOWLEDGEMENTS

This research has been funded by the Claude-Leon foundation. Some of the observations reported in this paper were obtained with the Southern African Large Telescope (SALT). A special thanks to Debra Fischer, Andrei Tokovinin and the Georgia State University team for retrieving some of the reduced CHIRON spectra from Yale University's SMARTS archive. We are grateful to the following people: Fred Marang and Francois van Wyk, who made most of the IR observations, and Karen Pollard, Dave Laney and Brian Carter who also contributed. We thank High Point University for providing the funds necessary to purchase observing time on the CTIO 1.5 m telescope through the SMARTS Consortium. This study has made use of fitting routines from the Institute voor Sterrenkunde python repository of KU Leuven. DK acknowledges the support of the Australian Research Council DECRA grant (95213534). PAW and SM acknowledge funding from the South African NRF. BNB and BJH acknowledge support from the US National Science Foundation (grants AST-1812874 and 1413660, respectively).

## DATA AVAILABILITY

The near-IR photometry and radial velocity data can be accessed online through CDS.

## REFERENCES

Akashi M., Soker N., 2013, MNRAS, 436, 1961
Bakker E. J., Lamers H. J. G. L. M., Waters L. B. F. M., Waelkens C., 1996, A&A, 310, 861

# APPENDIX A: VOIGT PROFILE FITS TO OI ABSORPTION LINES

This paper has been typeset from a T<sub>E</sub>X/L<sup>A</sup>T<sub>E</sub>X file prepared by the author.

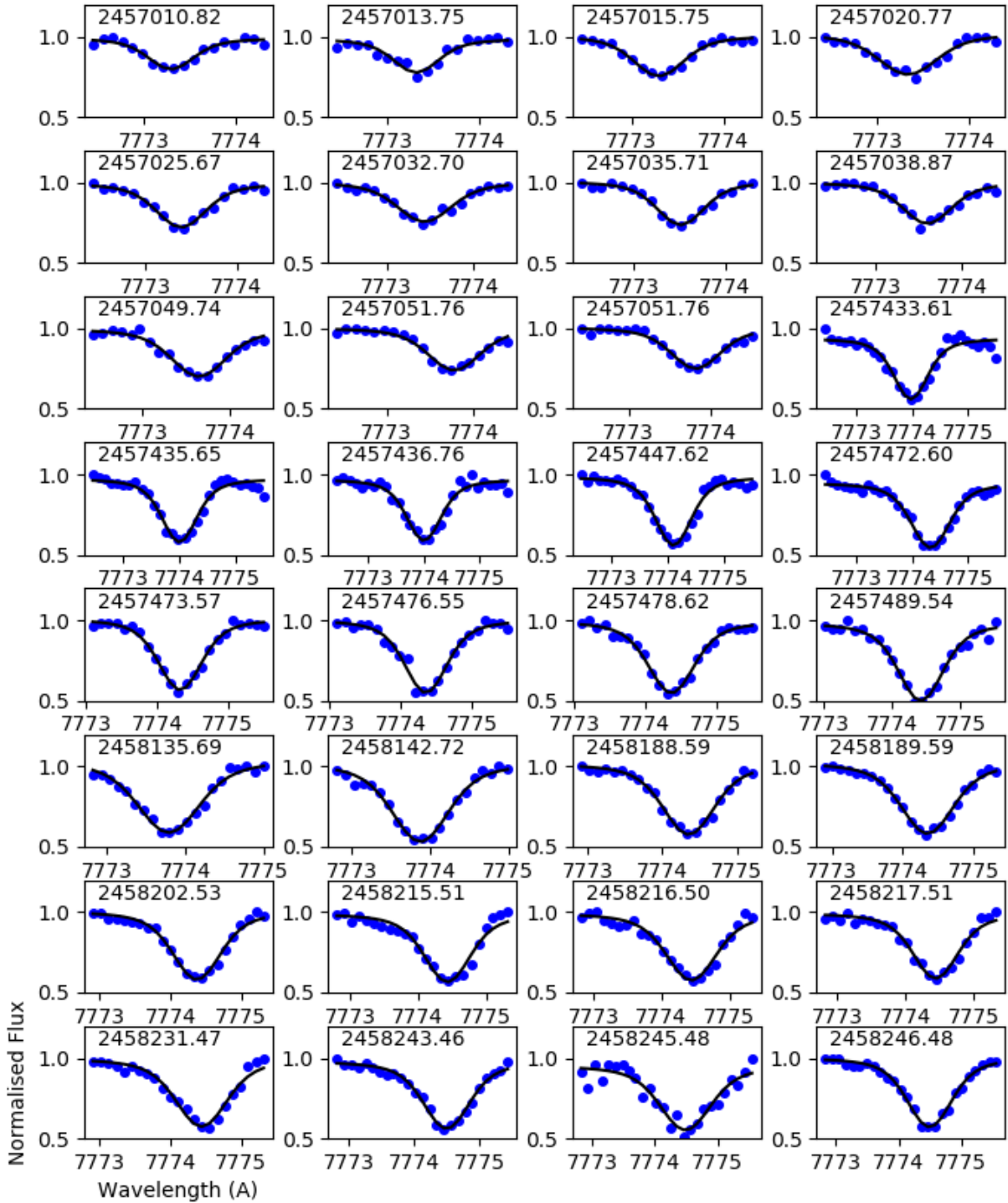

**Figure A1.** Voigt profile fits to the OI 7771 absorption line. The mid-exposure Julian dates are labeled on each plot.

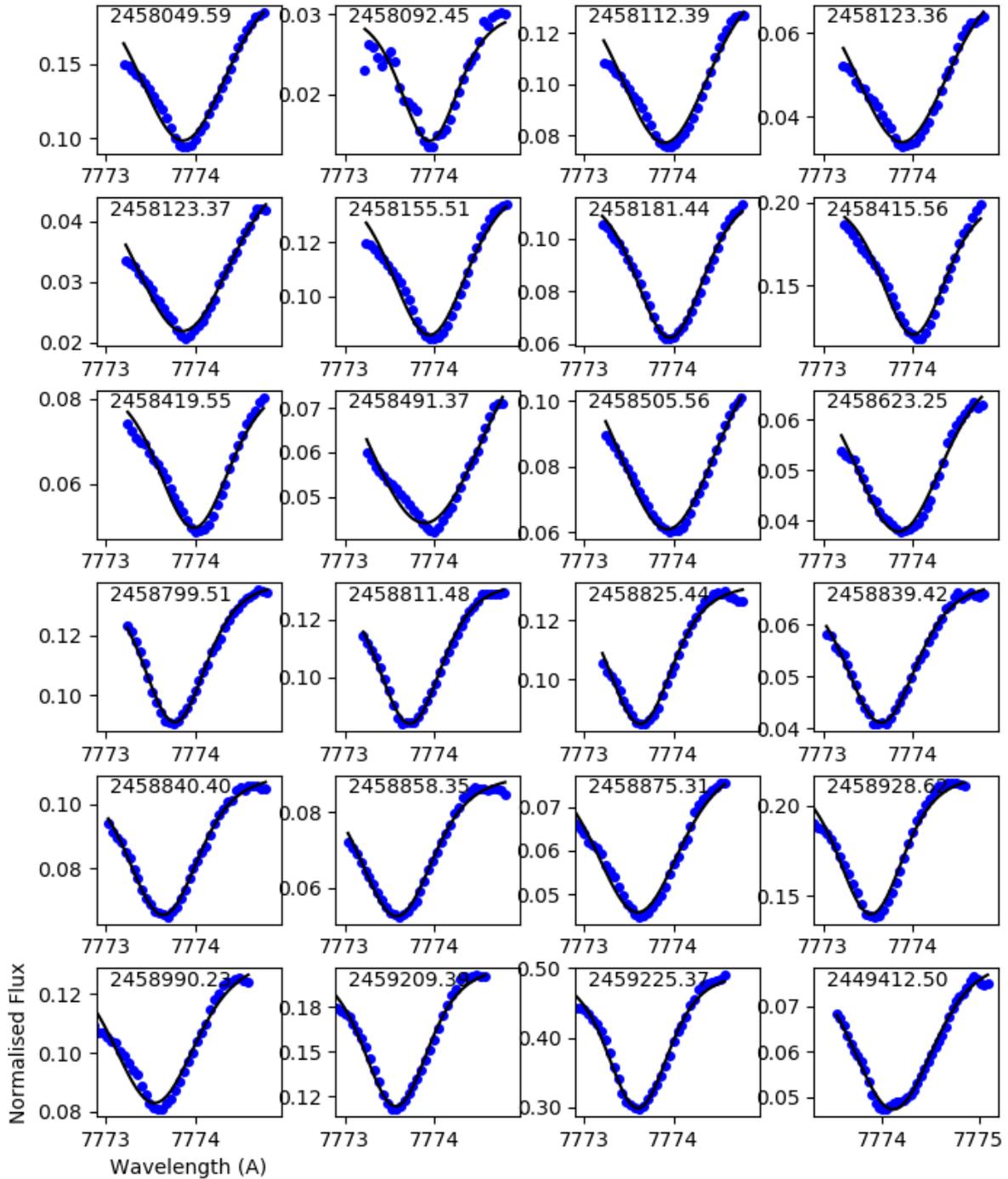

**Figure A2.** Voigt profile fits to the OI 7771 absorption line. The mid-exposure Julian dates are labeled on each plot.

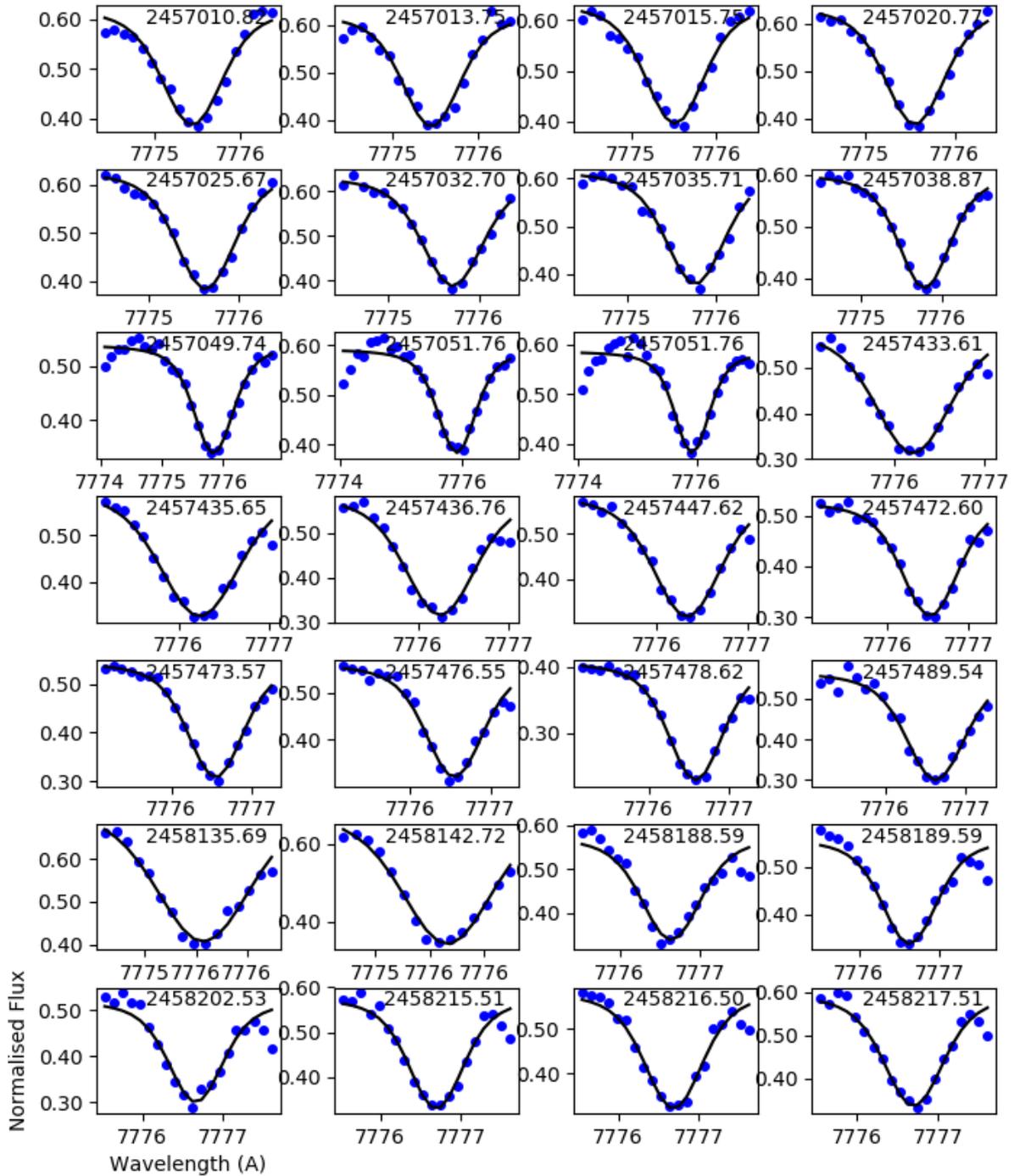

**Figure A3.** Voigt profile fits to the OI 7774 absorption line. The mid-exposure Julian dates are labeled on each plot.

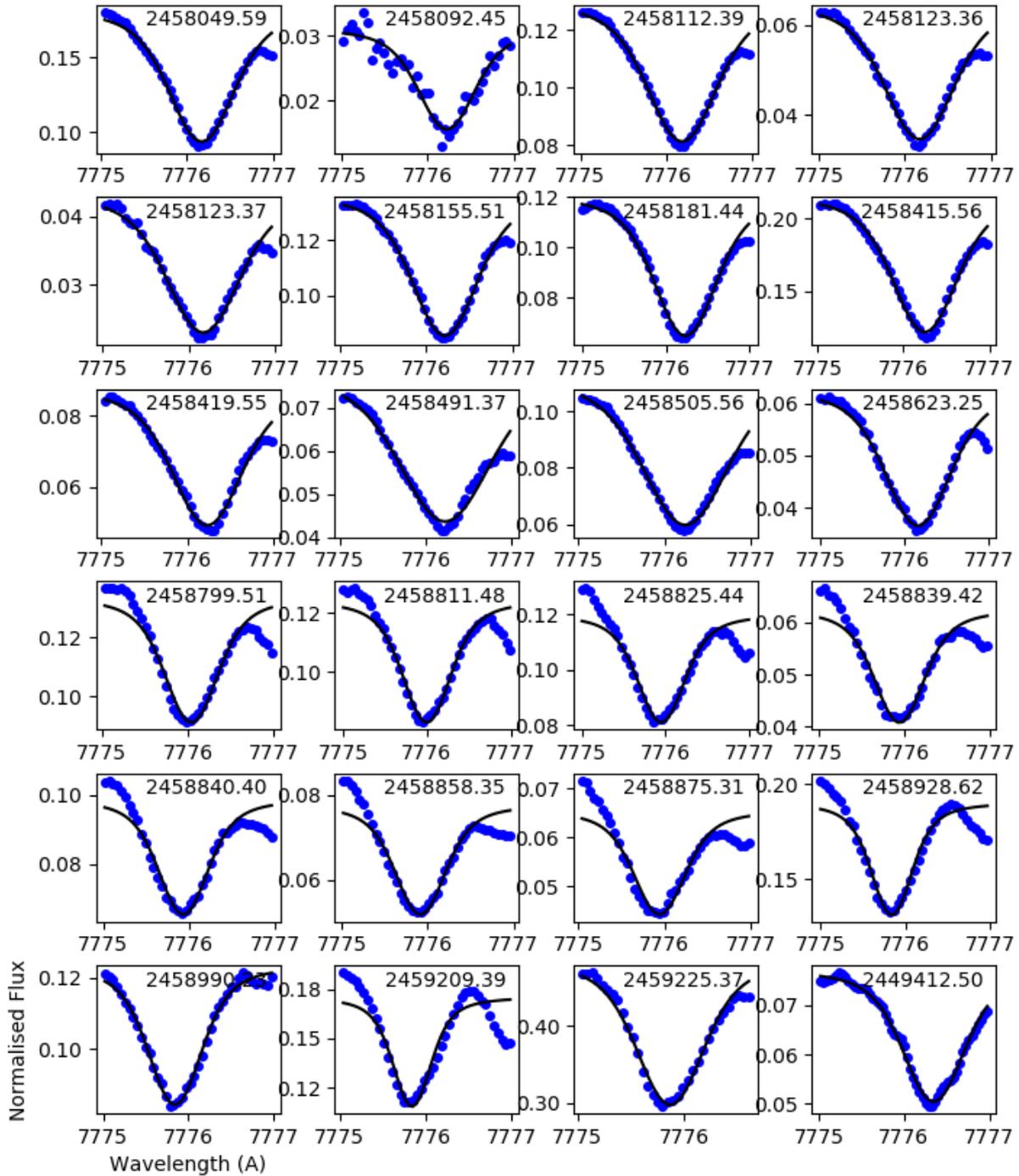

**Figure A4.** Voigt profile fits to the OI 7774 absorption line. The mid-exposure Julian dates are labeled on each plot.